\documentclass[preprint,12pt]{elsarticle}

\usepackage{amssymb}
\usepackage{hyperref}
\usepackage[x11names]{xcolor}
\usepackage{listings}
\usepackage{enumitem}
\usepackage{siunitx}
\usepackage{acronym}

\acrodef{ML}{machine learning}
\acrodef{RL}{reinforcement learning}
\acrodefindefinite{RL}{an}{a}
\acrodef{BO}{Bayesian optimization}
\acrodef{PS}{Proton Synchrotron}
\acrodef{LEIR}{Low-Energy Ion Ring}
\acrodef{SPS}{Super Proton Synchrotron}
\acrodef{LHC}{Large Hadron Collider}
\acrodef{FRS}{Fragment Separator}
\acrodef{TPC}{time projection chamber}

\journal{SoftwareX}

\begin{document}
\renewcommand{\labelenumii}{\arabic{enumi}.\arabic{enumii}}

\lstset{
    numbers=left,
    firstnumber=last,
    frame=lines,
    language=python,
    basicstyle=\footnotesize\ttfamily,
    commentstyle=\color{teal},
    stringstyle=\color{Firebrick3},
    classoffset=3,
    morekeywords={True,False,None},
    keywordstyle=\color{DeepPink3},
    classoffset=2,
    morekeywords={self},
    keywordstyle=\color{Blue4},
    classoffset=1,
    morekeywords={range,sum,map,filter},
    keywordstyle=\color{Green4},
    classoffset=0,
    morekeywords={as,yield},
    keywordstyle=\bfseries\color{Green4}
}

\begin{frontmatter}
 


\title{Geoff: The Generic Optimization Framework \& Frontend for Particle Accelerator Controls}


\author[GSI]{P.~Madysa}\ead{p.madysa@gsi.de}
\author[GSI]{S.~Appel}\ead{s.appel@gsi.de}
\author[CERN]{V.~Kain}\ead{verena.kain@cern.ch}
\author[CERN]{M.~Schenk}\ead{michael.schenk@cern.ch}
\address[GSI]{GSI Helmholtzzentrum für Schwerionenforschung, Darmstadt, Germany}
\address[CERN]{CERN, European Organization for Nuclear Research, Geneva, Switzerland}

\begin{abstract}
    Geoff is a collection of Python packages that form a framework for automation of particle accelerator controls.

    With particle accelerator laboratories around the world researching \acl{ML} techniques to improve accelerator performance and uptime, a multitude of approaches and algorithms have emerged.
    The purpose of Geoff is to harmonize these approaches and to minimize friction when comparing or migrating between them.
    It provides standardized interfaces for optimization problems, utility functions to speed up development, and a reference GUI application that ties everything together.

    Geoff is an open-source library developed at CERN and maintained and updated in collaboration between CERN and GSI as part of the EURO-LABS project.
    This paper gives an overview over Geoff's design, features, and current usage.
\end{abstract}

\end{frontmatter}



\section*{Metadata}
\label{metadata}

\begin{table}[!ht]
\begin{tabular}{|l|p{6.5cm}|p{6.5cm}|}
\hline
C1 & Current code version & 0.17.11 \\
\hline
C2 & Permanent link to code/repository used for this code version & \url{https://github.com/geoff-project/coi} \\
\hline
C3  & Permanent link to Reproducible Capsule & none \\
\hline
C4 & Legal Code License   & GPL-3.0-or-later OR EUPL-1.2+ \\
\hline
C5 & Code versioning system used & Git \\
\hline
C6 & Software code languages, tools, and services used & Gitlab, Python, PyTorch, Ruff \\
\hline
C7 & Compilation requirements, operating environments \& dependencies & Python 3.9+; NumPy, Gymnasium; the GUI application depends on CERN-internal components \\
\hline
C8 & If available Link to developer documentation/manual & \url{https://cernml-coi.docs.cern.ch/} \\
\hline
C9 & Support email for questions & \href{mailto:geoff-community@cern.ch}{\texttt{geoff-community@cern.ch}} \\
\hline
\end{tabular}
\caption{Code metadata}
\label{codeMetadata}
\end{table}

\section{Motivation and significance} 

%

The field of \textit{accelerator controls} is concerned with the development of hardware and software for the operation of a particle accelerator or a complex of accelerators.
It presents a number of challenges:
\begin{enumerate}[noitemsep]
    \item The hardware is operated \emph{around the clock} and subjected to considerable levels of radioactivity.
    \item \emph{Strong uptime requirements} mean that the software must be efficient, reliable and deterministic even in case of failures.
    \item Manpower, funds and \emph{upgrade efforts} are limited, even though accelerators operate for years and decades.
\end{enumerate}

Faced with this, software engineers responsible for building and maintaining controls systems used to eschew the latest technologies or dynamic programming languages, and favored more traditional and battle-tested technologies instead.
In practice, this means that middleware and control-room applications used the Java language and ecosystem.
This decision has traded performance for reliability\cite{Govindan_2023}.

At the same time, the field of \ac{ML} can no longer be ignored to satisfy the performance and uptime requirements with increasing accelerator complexity\cite{Edelen:2018jid}.
Early attempts to use \ac{ML} for accelerator controls\cite{Huang:2015wka,Appel:2019xlr} were successful in individual test cases, but didn't gain traction due to the limited \ac{ML} ecosystem in Java.
Most contemporary \ac{ML} research is based on the Python language and frameworks like TensorFlow and PyTorch.

While individual accelerator scientists have used these technologies on a case-by-case basis\cite{kain:2020vjs}, this has led to a lot of \textit{duplicated work}, \textit{lack of compatibility} and \textit{disregard for usability}.
This hampers adoption and makes comparison of the different approaches much more difficult\cite{ASRO_2023}.

\vspace{0.5\baselineskip}

Geoff\cite{madysa_2023_8434513} is a Python framework developed at CERN and open-sourced as part of the EURO-LABS\cite{EURO-LABS} project.
It provides:
\begin{itemize}[noitemsep]
    \item a \textit{standardized API} for numerical optimization and \ac{RL}, making it easier to share and compare algorithms;
    \item a \textit{library of utility functions} to avoid duplicate work;
    \item and a \textit{GUI application} to easily deploy solutions in the accelerator control room.
\end{itemize}

Geoff focuses on the execution of optimization algorithms and trained RL agents.
Tasks associated with the process of RL such as model storage and benchmarking are beyond its scope.

Because Geoff is written in Python\cite{Python:3.9}, the full ecosystem of the latest \ac{ML} is available.
In this way, it bridges the gaps between \ac{ML} experts, control-room operators, and accelerator physicists.

\vspace{0.5\baselineskip}

In a typical case, Geoff is deployed at a research institute by a group of software and IT infrastructure experts, e.g.\ the institute's controls department.
Operators---engineers managing the day-to-day operations of the accelerator from a control room---use the Geoff application to solve certain well-defined \textit{optimization problems}.
These problems are Python plugins written by domain experts with knowledge of specific accelerator subsystems---often physicists or engineers.

Geoff's open nature enables these groups to communicate and empowers operators to improve existing and create new plugins.
At the same time, its emphasis on versioning ensures that any erroneous updates can be easily rolled back.

\vspace{0.5\baselineskip}

Geoff is preceded by a number of similar frameworks\cite{Agapov:2014151,Piselli:ICALEPCS2017-TUPHA120,Geithner:IPAC2018-THPML028}.
It improves upon them with its radical approach to extensibility, making many problems solvable that would otherwise be impossible to express.

It competes\cite{roussel:2023yin} with Xopt\cite{roussel:ipac2023-thpl164} and Badger\cite{zhang:ipac2022-tupost058}.
Both have distinct advantages, but lack support for \ac{RL} or versioning and dependency management.

Geoff's API builds on that of the Gymnasium\cite{Towers:Gym2024} library.
By default, its application is bundled with state-of-the-art algorithms, such as provided by Stable-Baselines3\cite{Raffin:JMLR-v22-20-1364}, SciPy\cite{2020SciPy-NMeth} and Py-BOBYQA\cite{Cartis:10.1145/3338517,Cartis:10.1080/02331934.2021.1883015}.
It uses NumPy\cite{Harris:NumPy2020} for data transfer and PyQtGraph\cite{Moore:2023_8220855} for visualization.
Plugins can visualize their own data via Matplotlib\cite{Hunter:Matplotlib2007}.

Geoff is hosted at CERN and its DevOps automation is facilitated by CERN's Acc-Py project\cite{Elson:icalepcs2021-mopv040}.

\section{Software description} 


Geoff consists of a number of Python packages, some more central to its concepts than others.
The most important ones are as follows:
\begin{description}
    \item[cernml-coi]
        defines and standardizes the \textbf{C}ommon \textbf{O}ptimization \textbf{I}nterfaces interfaces for optimization problems.
        The interfaces are divided by capabilities so that an optimization problem can support numerical optimization, \ac{RL}, or both.
        Interfaces for \ac{RL} are provided by the Gymnasium package, which this package builds upon.

        Various mix-in interfaces can express additional capabilities, e.g.\ configurability via the GUI application.
        The package also provides routines that verify whether an implementation satisfies the invariants of these interfaces.

    \item[cernml-coi-optimizers]
        standardizes the interface specifically for numerical optimization algorithms.
        It also provides adapters to this interface for many popular libraries such as SciPy.
        The adapters are written as optional dependencies, so that they need not be installed if not required.

    \item[cernml-coi-utils]
        provides a number of utility functions and classes for optimization problems.
        These allow Geoff plugins to remain short and focused on the task at hand.
        Examples include a queue for simplified communications with an accelerator device and a method decorator to manage the state of a custom Matplotlib visualization.

    \item[geoff-app]
        is a reference implementation of a GUI application based on PyQt\cite{Qt:5.15} that loads a set of plugins and communicates with them via the above interfaces.
        Research institutes outside of CERN are encouraged to fork this project to take local concerns into account and provide accelerator-specific information.
        See section~\ref{sec:examples} for an example of how this is realized at CERN.
\end{description}

All of these packages follow Semantic Versioning\cite{SemVer} independently.
This allows them to evolve at different speeds and keeps version numbers meaningful;
e.g.\ a backwards-incompatible change to the utilities library (incrementing its MAJOR version number) has less impact than a similar change to the interfaces.

\subsection{Software architecture} 


\begin{figure}
    \begin{center}
        \includegraphics[height=0.2\textheight]{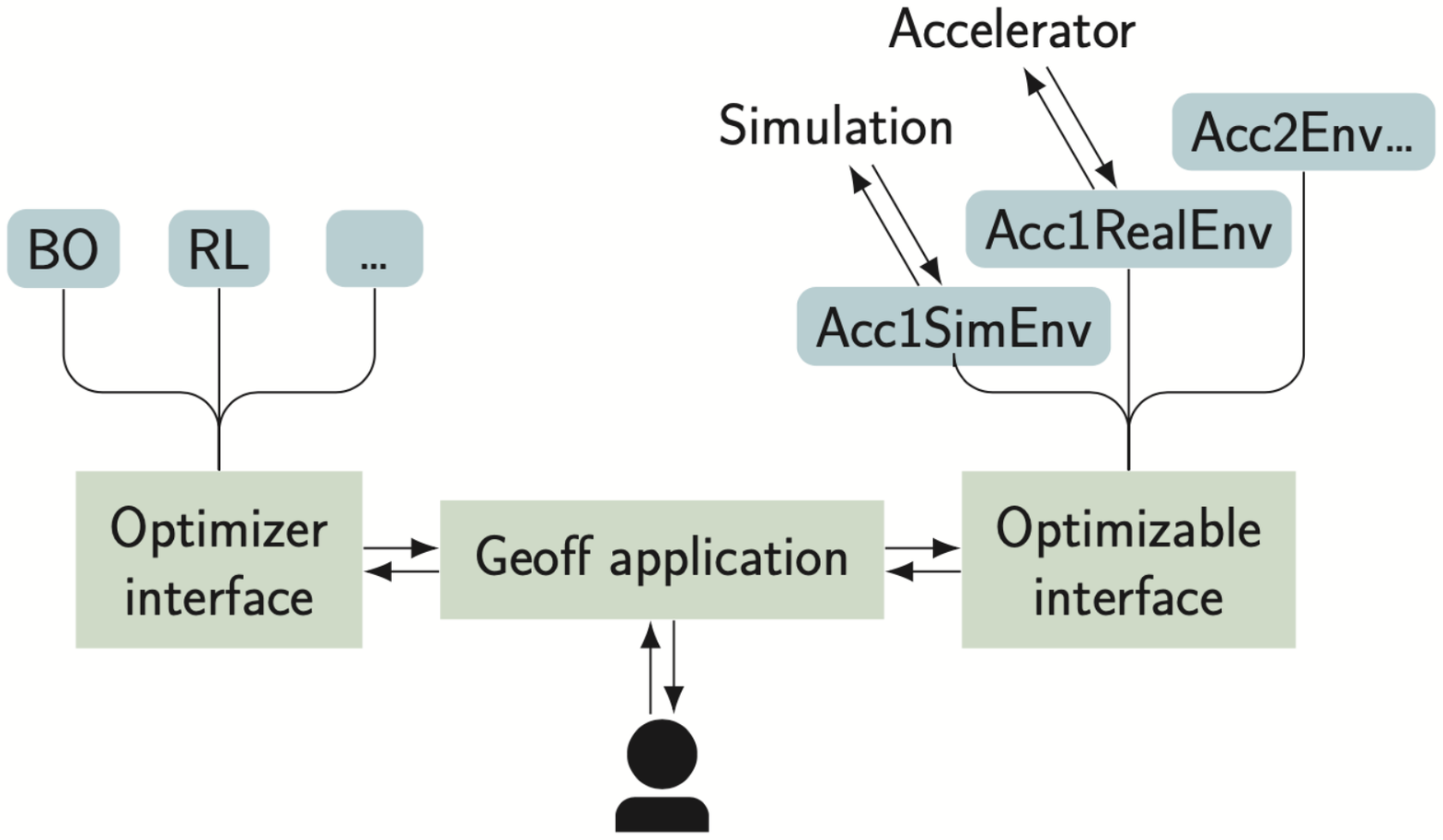}
    \end{center}
    \caption{Design architecture of \textit{Geoff}}\label{fig:architecture}
\end{figure}

The general architecture of the framework is shown in Fig.~\ref{fig:architecture}.
The user controls the system via a \textit{host application}.
This may be a GUI application like \textbf{geoff-app}, or a console script, or even a server endpoint.

The application loads available algorithms and optimization problems, either by direct import or via Python's \textit{entry points} API\cite{Python:EntryPoints}.
In addition to the \textit{generic} algorithms, \textit{custom} algorithms specific to a selected optimization problem may be loaded via \textit{custom-optimizer providers} (for numerical optimization) and \textit{custom-policy providers} (for \ac{RL}).

These providers are classes or functions that are each tied to one specific optimization problem and may define additional, custom-tailored algorithms.
This is useful in the case of \ac{BO}, where complicated kernel functions must often be defined in a way that makes sense only for a specific problem.
The host application can load these providers in two ways: either via an entry point, or via the optimization problem class if it implements the provider interfaces itself.
This makes it easy to publish additional algorithms without coordinating with whoever maintains the host application.

Once the user chooses an algorithm to use and a problem to solve, both are loaded dynamically via Python's import system, and the application communicates with them through the appropriate interfaces.
In particular, the application drives the interaction loop between optimization problem and algorithm, transmits data between them, and visualizes intermediate results to the user.

The optimization problem class, in turn, communicates with the controls system of the particle accelerator; i.e.\ it drives changed settings to the devices near the accelerator, and receives monitoring data from other devices.
Authentication of the user and authorization of the settings change are generally handled by the controls system itself\cite{Yastrebov:2010qnk}.
Geoff's architecture also allows the class to do anything \emph{besides} settings changes, including starting subprocesses and communicating with a simulation of the accelerator.

The interfaces defined by \textbf{cernml-coi-optimizers} supports \textit{closed-loop} implementations of optimization algorithms.
Such implementations typically only provide a \texttt{solve} function that, when called, executes the \emph{entire} algorithm and does not yield the control flow until the algorithm is completed.

In contrast, other frameworks like Xopt\cite{roussel:ipac2023-thpl164} require an \textit{open-loop} implementation.
Such implementations provide a state machine with methods sometimes called \texttt{ask} and \texttt{tell}.
When \texttt{ask} is called, it returns a candidate point $x$ to evaluate; \texttt{tell} is called with the value $f(x)$ of the objective function at that point.
Both methods should be called in alternation until the algorithm terminates.

The difference between open and closed loops is significant because most implementations of numerical optimization algorithms have a closed loop.
While it is trivial to translate an open loop to a closed loop, the reverse is usually \emph{impossible} without significantly rewriting the algorithm.
Admitting the more general case of closed-loop optimizers allows Geoff to support a wider range of algorithms than comparable frameworks.
BOBYQA\cite{Cartis:10.1145/3338517,Cartis:10.1080/02331934.2021.1883015} is one example of an algorithm for which no open-loop implementation has been published yet.

 \subsection{Software functionalities} 


 Geoff reduces the complexity of making $m$ optimization problems compatible with $n$ optimization algorithms from $\mathcal{O}(m\cdot{n})$ to $\mathcal{O}(m+n)$.
 The advantage is two-fold:
 For one, a newly developed algorithm can be used immediately on any existing problem, speeding up its adoption.
 For another, a given problem can be solved with a wide range of different algorithms, depending on situational context and evolving expertise of operators\cite{kain:ipac2024-tups55}.

 At all points, plugin developers have full freedom to use either the high-level abstractions of the Geoff utilities library, or to directly use the features of the underlying controls system.
 This allows plugins to solve not only simple toy problems, but also more complex ones, where e.g.\ an accelerator device is known to behave in an unusual fashion but it is not feasible to fix the issue at the source\cite{kain:ipac2024-tups55}.

 Because plugins are \emph{independent packages} with their own dependency declarations, they can scale from minimal proof-of-concept implementations to complex state machines that call out to subprocesses or request data from the accelerator's monitoring devices.
 Because plugins have their own versioning scheme, faulty upgrades are trivial to roll back without excessive downtime in the accelerator.

 The dynamic nature of the plugin architecture also allows plugin developers to test their code using a deployed version of the host application, and include it in a future one.
 The modular architecture of Geoff also means that plugin developers do not have to use the deployed application at all, and instead e.g.\ drive their plugin via a console script.
 This can reduce turnaround time in the early development cycle.

 \subsection{Sample code snippets analysis} 


This section shows an example implementation of a Geoff plugin, adopted from the developer documentation.
It adjusts a set of \textit{corrector magnets} in a section of the AWAKE accelerator at CERN\cite{AWAKE:2018gdq,PhysRevAccelBeams.23.124801}.
An electron beam traveling through the beam line is measured via a set of \textit{beam position monitors}.
The objective of the optimization problem is to adjust the correctors such that the beam travels as closely to the central axis of the monitors as possible.

The first lines import a number of packages (Gymnasium\cite{Towers:Gym2024}, NumPy\cite{Harris:NumPy2020}, Matplotlib\cite{Hunter:Matplotlib2007}) and the \textit{Common Optimization Interfaces} of Geoff.
PyJapc is a CERN-internal package to communicate with the controls system.

\begin{lstlisting}[lastline=8]
import gymnasium as gym
import numpy as np
from cernml import coi
from gymnasium.spaces import Box
from matplotlib import pyplot as plt
from pyjapc import PyJapc

\end{lstlisting}

Geoff represents optimization problems as Python classes.
They inherit from one or more of the standard interfaces, depending on whether they support single-objective numerical optimization (represented by the abstract base class \texttt{SingleOptimizable}), \ac{RL} (represented by \texttt{Env}), or both.

\begin{lstlisting}[linerange=10-10]
class AwakeSteering(coi.SingleOptimizable, gym.Env):
\end{lstlisting}

All classes contain a class-scope dictionary called \texttt{metadata}.
It contains information that a host application can use to infer how to interact with the plugin.
Some dictionary keys are standardized by Gymnasium and some by Geoff.
Geoff's developer documentation describes all standard keys.
Laboratories are also encouraged to standardize their own, appropriately prefixed keys.
All keys are considered optional and have reasonable default values.

\begin{lstlisting}[linerange=11-16]
    metadata = {
        "render_modes": ["human"],
        "cern.machine": coi.Machine.AWAKE,
        "cern.japc": True,
    }
\end{lstlisting}

Here, the list behind \texttt{render\_modes} specifies the ways in which the plugin state can be visualized.
Gymnasium defines the mode \texttt{"human"} for interactive visualization, typically initiated without a host application.
We implement this via Matplotlib, but other libraries can be used.
The other keys are laboratory-specific:
\texttt{"cern.machine"} marks the relevant accelerator in the CERN complex.
CERN's host application filters the optimization problems presented to the user based on this value.
If \texttt{"cern.japc"} is set, CERN's host application handles user authentication for the plugin and passes the configured \texttt{PyJapc} object to its \texttt{\_\_init\_\_} method.

\begin{lstlisting}[linerange=17-19]
    CORRECTORS = ["logical.RCIBH.430029/K", ...]
    MONITORS = ["TT43.BPM.430028/Acquisition", ...]
\end{lstlisting}

These lines are not part of the Geoff interface.
They list the \textit{device names} that will be passed to PyJapc to communicate with the correct devices on the accelerator.
They have been abbreviated here for brevity.

The following lines define the mathematical \textit{domain} of the algorithm: this is \texttt{optimization\_space} for numerical optimization and \texttt{action\_space} for \ac{RL}.
\ac{RL} algorithms additionally require the domain of possible observations for normalization purposes (\texttt{observation\_space}).

\begin{lstlisting}[linerange=20-23]
    action_space = Box(-0.1, 0.1, shape=[len(CORRECTORS)])
    observation_space = Box(-1.0, 1.0, [2 * len(MONITORS)])
    optimization_space = Box(-1.0, 1.0, [len(CORRECTORS)])
\end{lstlisting}

Domains must be subclasses of \texttt{gymnasium.Space}.
We use \texttt{Box} here, which represents a multidimensional rectangular bounding box.
While \texttt{Box} is the most common type of space, discrete and composite spaces exist as well.

Next, an initializer stores the chosen render mode and uses private helper methods to fetch the state of the machine from the controls system:

\begin{lstlisting}[linerange=24-29]
    def __init__(self, japc=None, render_mode=None):
        self.render_mode = render_mode
        self.japc = japc or PyJapc()
        self.latest_readings = self._recv_bpm_readings()
        self.correctors = self._recv_corrector_values()
\end{lstlisting}

These helper methods are defined directly below.
They are the only part of the class that directly interacts with the controls system:

\begin{lstlisting}[linerange=30-45]
    def _recv_bpm_readings(self):
        props = [self.japc.getParam(m) for m in self.MONITORS]
        return np.array(
            [[p["horPos"], p["verPos"]] for p in props]
        ).flatten()

    def _recv_corrector_values(self):
        return np.array([
            self.japc.getParam(c) for c in self.CORRECTORS
        ])

    def _send_corrector_values(self, values):
        self.correctors = values.copy()
        for addr, value in zip(self.CORRECTORS, values):
            self.japc.setParam(addr, value)
\end{lstlisting}

Next comes the implementation of the interface methods for \ac{RL} and numerical optimization.
Both interfaces require two methods.
First, a method that initializes either \iac{RL} episode or an optimization run (\texttt{reset} and \texttt{get\_initial\_params} resp.):

\begin{lstlisting}[linerange=46-57]
    def reset(self, *, seed=None, options=None):
        super().reset(seed=seed, options=options)
        correctors = self.optimization_space.sample()
        self._send_corrector_values(correctors)
        info = {}
        if self.render_mode == "human": self.render()
        return self.correctors.copy(), info

    def get_initial_params(self, *, seed=None, options=None):
        params, _info = self.reset(seed=seed, options=options)
        return params
\end{lstlisting}

Secondly, a method that performs a step of the algorithm and returns new data (\texttt{compute\_single\_objective} for optimization, \texttt{step} for \ac{RL}):

\begin{lstlisting}[linerange=58-71]
    def compute_single_objective(self, params):
        self._send_corrector_values(params)
        self.latest_readings = self._recv_bpm_readings()
        if self.render_mode == "human": self.render()
        return np.sqrt(np.mean(self.latest_readings**2))

    def step(self, action):
        correctors = self.correctors + action
        reward = -self.compute_single_objective(correctors)
        obs = self.latest_readings.copy()
        terminated = reward > -0.1
        truncated = False
        info = {}
        return obs, reward, terminated, truncated, info
\end{lstlisting}

All methods \emph{also} perform a rendering step \emph{if} the render mode is \texttt{"human"}.
This is required by Gymnasium's definition of the render mode.

Both interfaces define default attributes and methods that can be overridden for customization.
For example, \texttt{param\_names} allows \texttt{SingleOptimizable} to attach names to the individual axes of its \texttt{optimization\_space}:

\begin{lstlisting}[linerange=73-74]
    param_names = CORRECTORS
\end{lstlisting}

Another example is the \texttt{render} method, which implements visualization:

\begin{lstlisting}[linerange=75-87]
    def render(self):
        if self.render_mode != "human":
            return super().render()
        figure = plt.figure("AwakeElectronBeamSteering")
        figure.clear()
        plt.plot(self.latest_readings, "|-")
        plt.xlabel("BPM")
        plt.ylabel("Beam position (mm)")
        figure.show(warn=False)
        figure.canvas.draw_idle()
        return None

\end{lstlisting}

Finally, the plugin registers itself with Geoff:

\begin{lstlisting}[linerange=88-90]
coi.register("AwakeSteering-v1", entry_point=AwakeSteering)

\end{lstlisting}

This ensures that the plugin will be found when a host application aggregates all available plugins.
While registration occurs here at runtime, Geoff also supports entry points for registration at installation.

\section{Illustrative examples} 
\label{sec:examples}

%
%

\begin{figure}
    \begin{center}
        \includegraphics[height=0.3\textheight]{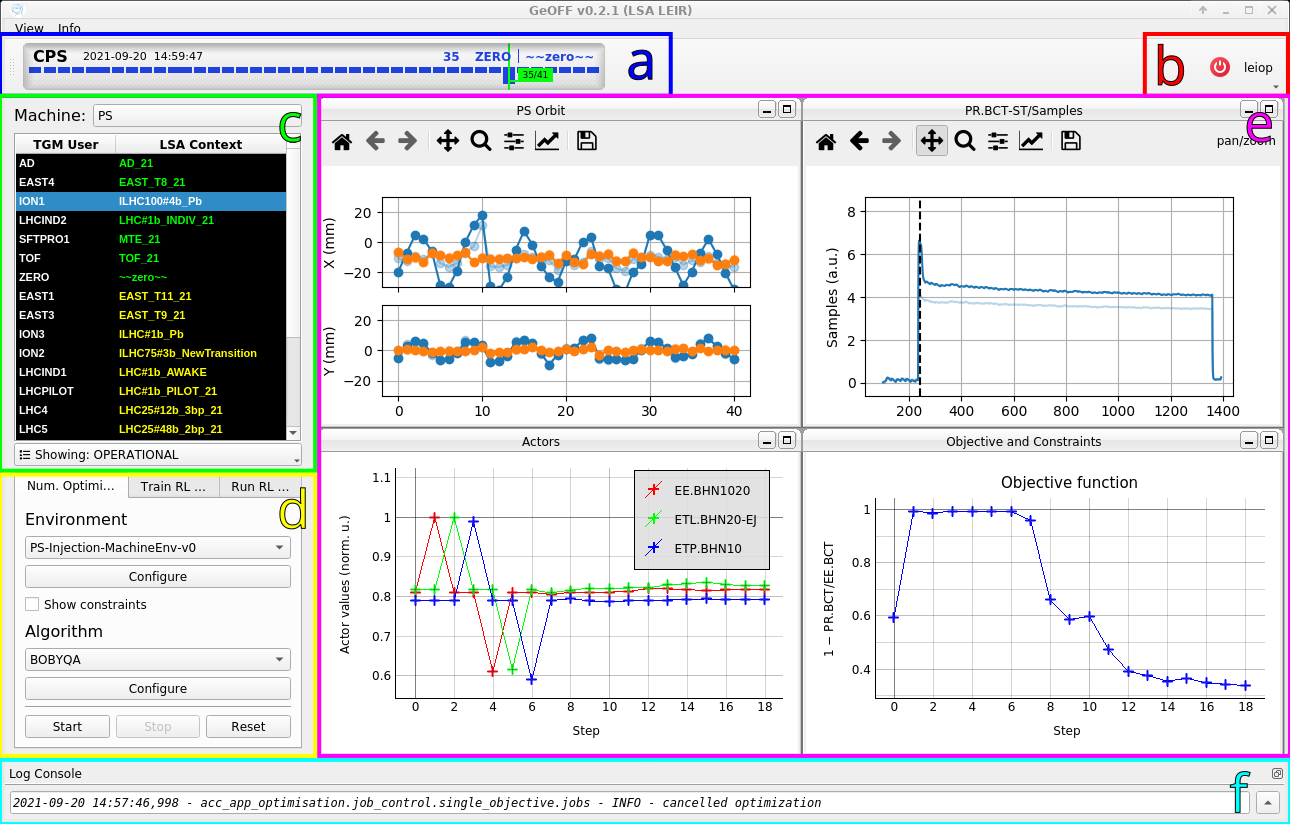}
    \end{center}
    \caption{%
        Example of the Geoff GUI application deployed in the CERN control room.
        The various GUI components are framed in several colors and labeled a--f.
        See the text for their description.
    }\label{fig:example}
\end{figure}

Figure~\ref{fig:example} shows a screenshot of the Geoff GUI application that is deployed in the CERN control room.
It shows the application state after a successful run of the BOBYQA algorithm.

The optimization problem here was to maximize the intensity of the particle beam injected into the \ac{PS}, a particle accelerator at CERN.
The main difficulty lies in the communication with devices in different \textit{timing domains} (i.e.\ with unaligned data publishing rates).
Data from these devices must potentially be polled multiple times per iteration of the optimization algorithm and carefully synchronized.
This would be impossible to implement in a framework where data acquisition is the framework's responsibility rather than the plugin's.

The central area of the window \colorbox{magenta}{\color{white}(e)} is taken up by four sub-windows that show live updates during optimization:
\begin{itemize}[noitemsep]
    \item The top windows show the status of the most recent step; they are provided by the plugin itself.
    \item The bottom windows show the progress of optimization; they are always provided by the host application.
\end{itemize}

The optimization control area \colorbox{yellow}{(d)} allows the user to choose between numerical optimization, training an \ac{RL} agent and executing an \ac{RL} agent that has already been trained.
The user can also start, cancel and revert these procedures here.
This area is also where the user selects both the problem to solve and the algorithm to use.

The application window shown here also contains CERN-specific GUI elements such as:
\begin{itemize}[noitemsep]
    \item the current schedule of acceleration cycles \colorbox{blue}{\color{white}(a)};
    \item a log-in button that may be required when modifying mission-critical settings \colorbox{red}{\color{white}(b)};
    \item a selector for the acceleration cycle whose settings are modified \colorbox{green}{(c)};
    \item an expandable status line with a log of important events \colorbox{cyan}{(f)}.
\end{itemize}
These elements are maintained by the CERN-internal Acc-Py project\cite{Elson:icalepcs2021-mopv040}.
Other research institutes are expected to develop GUI elements that are relevant to their respective operations.

\begin{figure}
    \begin{center}
        \includegraphics[height=0.25\textheight]{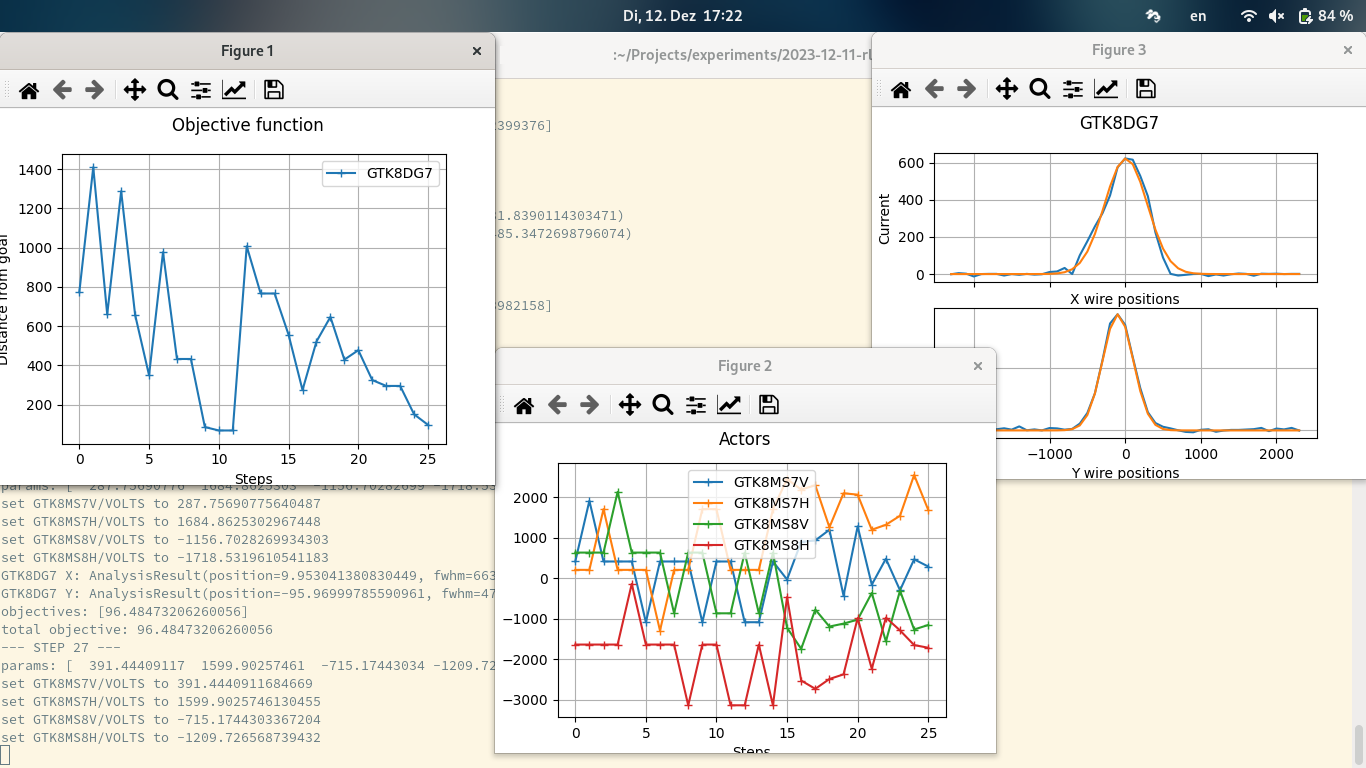}
    \end{center}
    \caption{%
        Example of Geoff being run in a terminal in the GSI control room
    }\label{fig:terminal}
\end{figure}

While Geoff supports a GUI host application, it doesn't require one.
As an example, when porting efforts at GSI began, no GUI application without CERN dependencies existed.
Nonetheless, the framework could be used on a command-line terminal.
Figure~\ref{fig:terminal} shows a screenshot of such an optimization run at GSI.
Logging messages were being printed in the terminal, while the optimization progress was visualized via Matplotlib and X11 window forwarding.

\section{Impact} 

%
%

Geoff has enabled studies where a comparison of numerical optimization and \ac{RL} is of interest\cite{madysa:ipac2022-tupost040} and where the complex optimization procedures excluded less flexible frameworks\cite{bruchon:icalepcs2023-tupdp086}.

It has improved the pursuit of other research questions by
making it easier to share optimization algorithms\cite{uden:ipac2023-tupa159};
facilitating the migration to better algorithms\cite{kain:ipac2024-tups55};
and reducing the time from proof of concept to operational deployment\cite{fraser:ipac2022-wepotk043}.

It is used extensively in the daily operations of the LINAC4 linear accelerator and the \ac{PS} Booster\cite{skowronski:ipac2022-tupost041},
the \ac{PS}\cite{huschauer:ipac2022-mopost006,huschauer:ipac2023-tupa158},
and the \ac{LEIR}\cite{biancacci:ipac2022-wepopt055,AlemanyFernandez:2872882,alemany-fernandez:ipac2023-tupa151}.
\Ac{LEIR} particularly benefits from Geoff because it has no dedicated operators and most tasks used to be done manually.

\begin{figure}
    \begin{center}
        \includegraphics[height=0.2\textheight]{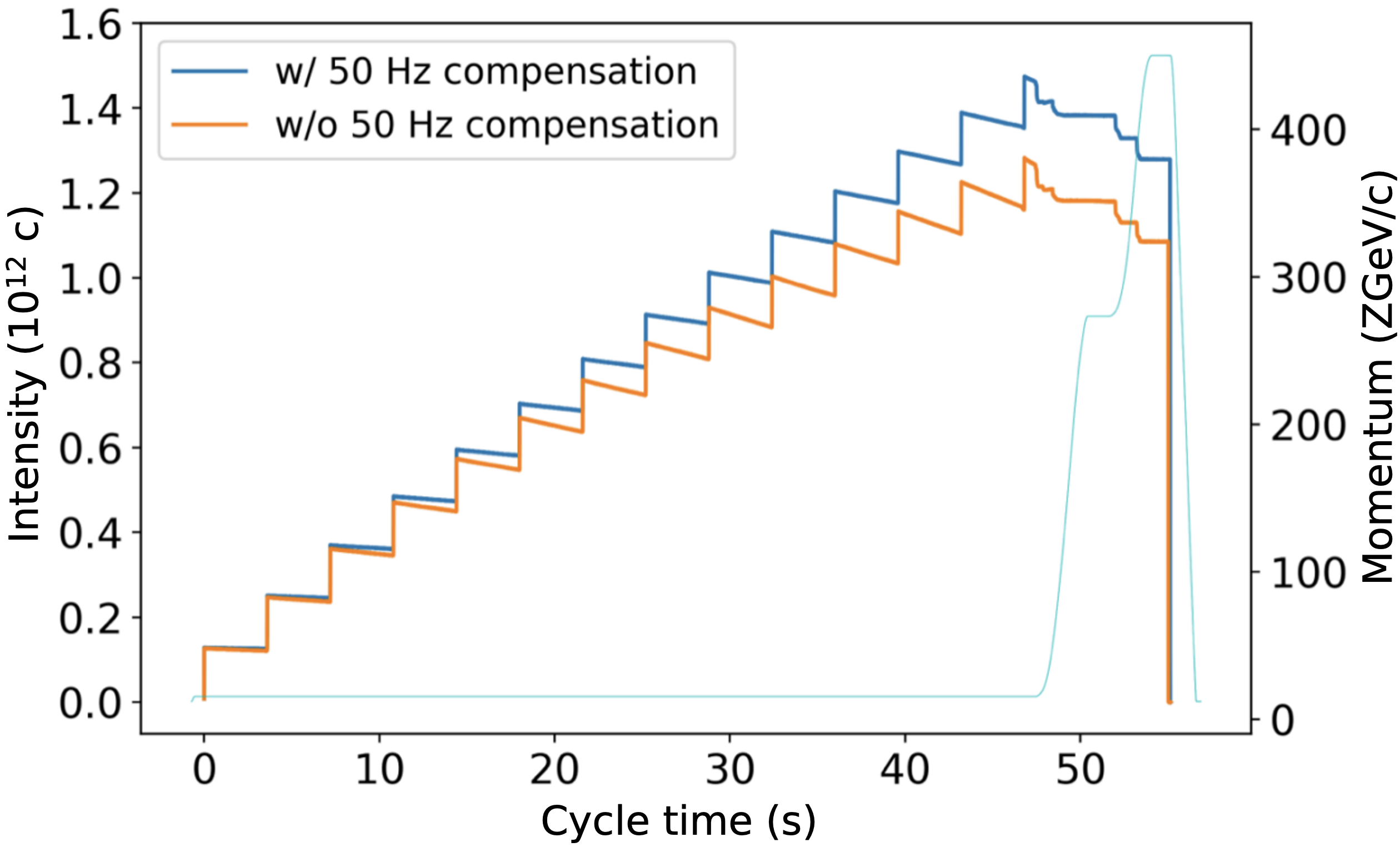}
    \end{center}
    \caption{%
        Intensity of the beam injected from the \acs*{SPS} into the \acs*{LHC} with and without the \acl*{BO} of main-power converter noise\protect\cite{kain:malapa-2025}
    }\label{fig:transmission_ions}
\end{figure}

At the \ac{SPS}, Geoff reduced the positioning time of the slow-extraction septa from eight hours to ten minutes\cite[p.~9]{madysa:gcm2022-ml4acc}.
It was also used in the compensation of main-power converter noise from the electrical grid and allowed the continuous upgrade of algorithms over multiple years;
the latest one (based on \ac{BO}), improves ion beam transmission to the \ac{LHC} by \qtyrange[range-units=single, range-phrase=\text{--}]{15}{20}{\percent}\cite{kain:malapa-2025}, as Fig.~\ref{fig:transmission_ions} shows.
Finally, Geoff facilitated the operational deployment of simulation-trained \ac{RL} agents for beam trajectory correction in the transfer lines from \ac{SPS} to the North Experimental Area\cite{menor:rl4aa-2025}.

At the source of ion beams for the LINAC3 linear accelerator, Geoff is used to optimize and continuously tune various parameters to provide a high and stable output current for ion operations\cite{kuchler:ecris2024-mop11}.
The problem is a mix of slow and fast dynamics and the nested-loop structure used to solve it would be difficult to implement without Geoff's flexibility.

Geoff is used at every accelerator of the CERN complex except the \ac{LHC}\cite{madysa:gcm2022-ml4acc,kain:ipac2022-weiygd1,trad:japw-2024}, which was designed from the start in a way that avoids black-box optimization and instead prefers non-iterative, model-based algorithms.

\vspace{0.5\baselineskip}

\begin{figure}
    \begin{center}
        \includegraphics[height=0.25\textheight]{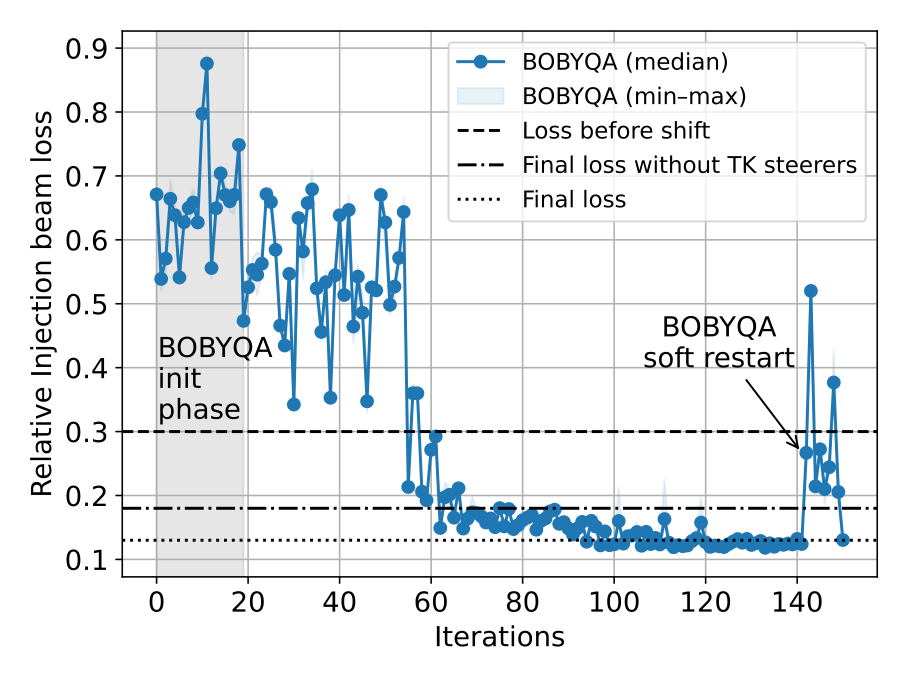}
    \end{center}
    \caption{%
        Automatic online steering of the SIS18 multi-turn injection using the BOBYQA optimization algorithm\protect\cite{appel:ipac2024-mops68}.
        The gray shaded area marks the initialization phase of the algorithm.
        The final point presents an additional evaluation at the found optimum; it is not strictly part of the algorithm.
    }\label{fig:loss_sis18}
\end{figure}

Geoff is also being deployed at GSI\cite{roussel:2023yin}.
Successful tests have been performed in simulation\cite{appel:ipac2024-tups59} and on real accelerators.
The beam loss at injection into the SIS18 synchrotron could be reduced from \qtyrange{40}{15}{\percent}, as Fig.~\ref{fig:loss_sis18} shows\cite{appel:ipac2024-mops68}.
At GSI's \ac{FRS}, Geoff has been used to adjust the ion beam trajectory.
Its flexibility was crucial due to the large number of interacting components\cite{appel:icap2024}.

Finally, as part of EURO-LABS\cite{EURO-LABS}, efforts are ongoing to use Geoff at the laser facility at CEA Paris-Saclay.
Preliminary results are expected at the end of 2025.

Commercial use of Geoff is currently not foreseen, given its large focus on an academic context.
However, such use is possible and permitted by its open-source license.

\section{Conclusions} 

Geoff successfully unified previously uncoordinated efforts to use \ac{ML} for accelerator control automation at CERN.
It is used throughout the CERN complex; this ensures maintenance and further development for years to come.
At the same time, its support by EURO-LABS made it possible to make it open-source and available to other research institutes.

Future plans for the software are outlined in a roadmap in the developer documentation (see table~\ref{codeMetadata}).
One such plan is the inclusion of support for multi-objective optimization\cite{Abdulrahman_Niu_2023} via a framework like PyMoo\cite{pymoo}.
Support for this branch of optimization would greatly expand the scope in which Geoff can be used.
Another plan is to provide tools to allow bootstrapping a Bayesian optimizer with previously recorded data to greatly speed up experimentation with such algorithms.
Finally, we intend to provide a redesigned reference GUI application with support for data collection and export to encourage use by non-experts in a wider range of accelerator laboratories.

\section*{Acknowledgements} 
\label{acknowledgments}

We thank our CERN colleagues for their help in improving and extending Geoff.
This includes Francisco Huhn, Ivan Sinkarenko, Michi Hostettler, Niky Bruchon, and Philip Elson.

We further thank our colleagues from the Super-FRS at GSI for their help in testing Geoff in a new environment:
Daniel Kallendorf, Erika Kazantseva, Helmut Weick, Martin Bajzek, and Stephane Pietri.

We also thank our colleagues who added Python support to the GSI controls system: Adrian Oeftiger, Dominic Day, Jutta Fitzek, Ralph Baer, Raphael Mueller, and Udo Krause.

The EURO-LABS project has received funding from the European Union's Horizon Europe Research and Innovation programme under Grant Agreement no.~\num{101057511}.




\bibliographystyle{elsarticle-num}
\bibliography{references}


%
%

\end{document}